\documentclass[aps,prb,twocolumn,groupedaddress]{revtex4}

\usepackage[dvips]{color}
\usepackage{graphicx,color}
\begin{document}

\title{Quantum Dot in 2D Topological Insulator: The Two-channel Kondo Fixed Point }

\author{ K. T. Law$^{1,2}$, C. Y. Seng$^3$, Patrick A. Lee$^{2}$, and T. K. Ng$^3$}

\affiliation{$^1$ Institute for Advanced Study, Hong Kong University of Science and Technology, Hong Kong, People's Republic of China\\
$^2$Department of Physics, Massachusetts Institute of Technology, Cambridge, Massachusetts, 02139, USA\\
$^3$Department of Physics, Hong Kong University of Science and Technology, Hong Kong, People's Republic of China}

\begin{abstract}

In this work, a quantum dot couples to two helical edge states of a 2D topological insulator through weak tunnelings is studied. We show that if the electron interactions on the edge states are repulsive, with Luttinger liquid parameter $ K < 1 $, the system flows to a stable two-channel fixed point at low temperatures. This is in contrast to the case of a quantum dot couples to two Luttinger liquid leads. In the latter case, a strong electron-electron repulsion is needed, with $ K<1/2 $, to reach the two-channel fixed point. This two-channel fixed point is described by a boundary Sine-Gordon Hamiltonian with a $K$ dependent boundary term. The impurity entropy at zero temperature is shown to be $ \ln\sqrt{2K} $. The impurity specific heat is $C \propto T^{\frac{2}{K}-2}$ when $ 2/3 < K < 1 $, and $ C \propto T$ when $ K<2/3$.  We also show that the linear conductance across the two helical edges has non-trivial temperature dependence as a result of the renormalization group flow.
 
\end{abstract}


\maketitle

\emph{Introduction}---The two-channel Kondo model has been under intense theoretical study\cite{CZ} after the seminal work of Noziere and Blandin\cite{NB}, in which they pointed out that a non-Fermi liquid fixed-point exists for the two-channel Kondo model with non-interacting electrons. Recently, the authors of Ref.\onlinecite{PRSOG} demonstrated experimentally that a quantum dot couples to an infinite reservoir with non-interacting electrons and a finite reservoir with interacting electrons may display two-channel Kondo effect. The key idea is that Coulomb blockade suppresses the exchange of electrons between the finite reservoir and the infinite reservoir.\cite{OG} Consequently, the two reservoirs couple to the quantum dot independently. The two-channel Kondo quantum critical point can be reached by tuning the couplings of the quantum dot with the reservoirs to equal strength.

Indeed, using electron-electron repulsion to suppress interreservoir tunnelings was suggested earlier by Fabrizio and Gogolin, in the context of a quantum dot couples to two Luttinger liquid leads.\cite{OG,FG} They show that the two-channel Kondo fixed point can be reached when the repulsive interactions of the electrons in the leads are strong enough, i.e. with Luttinger liquid parameter $K < 1/2$. Similar conclusions were also made in later works.\cite{Kim,CRS} However, two-channel Kondo effect in Luttinger liquids have never been observed experimentally.

Recently, a new class of materials called topological insulators were first theoretically proposed and then experimentally fabricated.\cite{KM,BHZ,FKM,Koniq,Hsieh} Two dimensional topological insulators have gapless helical edge states despite the presence of a bulk gap.\cite{KM} The edge states are called helical because the directions of the spins and the momentum of the electrons are tied together.\cite{WBZ} The one-channel Kondo effect of helical edge states was first studied in Ref.\onlinecite{SI} as a truncated model of Kondo effect in Luttinger liquids. A more detailed recent study is carried out in Ref.\onlinecite{MLOQWZ}.
\begin{figure}
\includegraphics[width=3.4in]{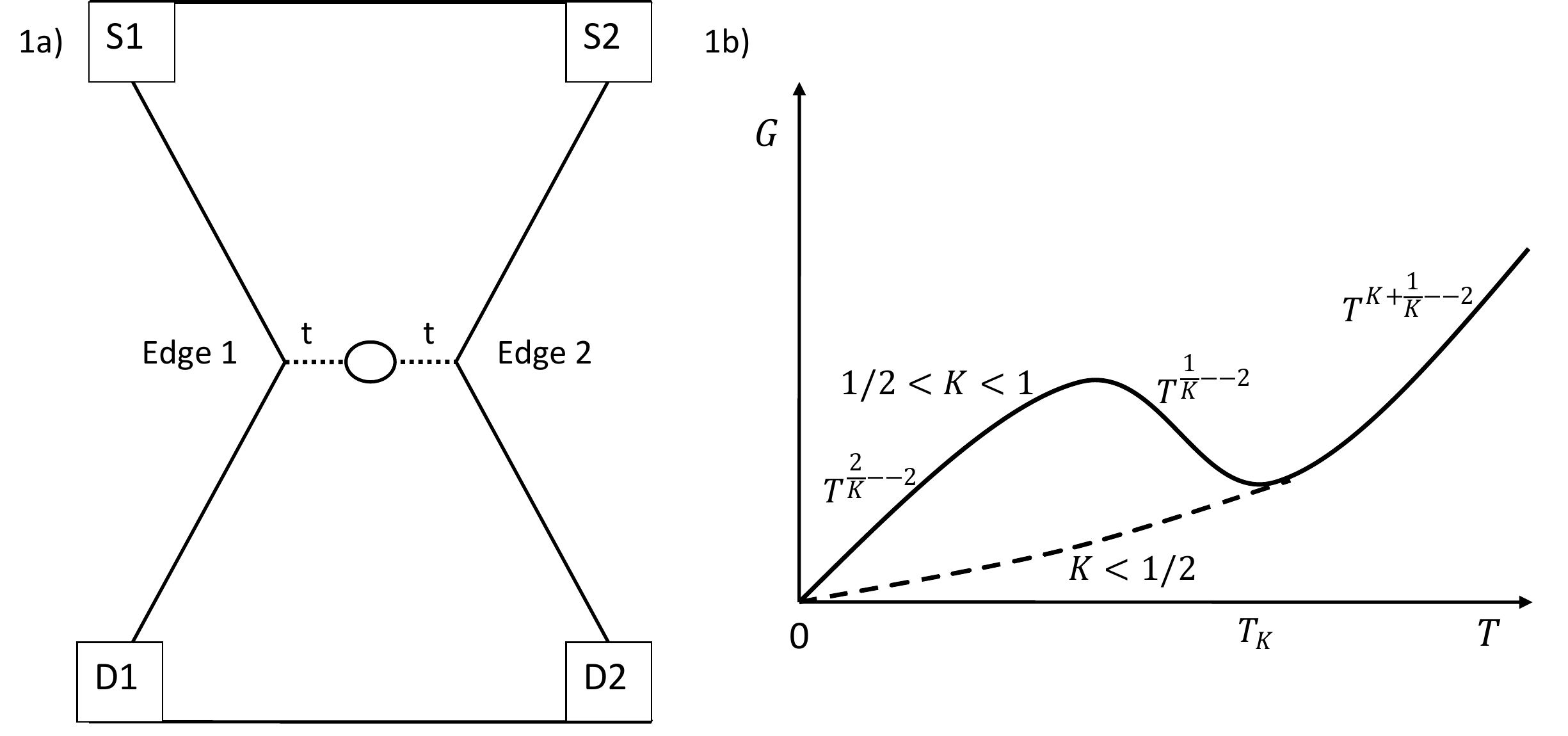}
\caption{\label{edge} a) A quantum dot couples to two helical edges of a 2D topological insulator through electron tunneling. S and D denote the source and drain respectively. b)The schematic picture of the linear conductance $G$ from S1 to D2 versus temperature $T$. The solid line and the dashed line depict the $1/2<K<1$ and $K<1/2$ cases respectively.}
\end{figure}

In this work, we study a quantum dot couples to two helical edge states as depicted in Fig.1a. If the electrons on the edge states are non-interacting, it is well known that the system can be described by a one-channel Kondo Hamiltonian.\cite{GR,NL} However, we show that a weak repulsive interaction, with $ K<1 $ is enough to drive the system to the two-channel Kondo fixed point. This is in sharp contrast to the case of Luttinger liquid leads in which $K <1/2$ is needed.\cite{FG} This two-channel fixed point is described by a boundary Sine-Gordon Hamiltonian with a $K$ dependent boundary term. We study the ground state, thermaldynamic and transport properties near this two-channel fixed point. More specifically, we show that the impurity entropy at zero temperature is $ \ln\sqrt{2K} $, the impurity specific heat $C_{imp} \propto T^{\frac{2}{K}-2}$ when $ 2/3 < K < 1 $, and $ C_{imp} \propto T$ when $ K<2/3$, and the linear conductance across the two helical edges through the quantum dot has non-trivial temperature dependence as a result of the renormalization group flow.

\emph{The Model}---The schematic diagram of our model is depicted in Fig.1a. Two helical edge states of an 2D topological insulator are brought close to each other at a tunneling junction. A quantum dot is placed at the middle of the junction and couples to the two edge states through electron tunnelings. The system is described by an Anderson Hamiltonian 
$ H_{A} = H_{0} + H_{i} + H_{d} + H_{t} $ where

$ H_{0}= \sum\limits_{i=1,2}-iv_{F}\int[\psi_{iR\uparrow}^{\dagger}(x)\partial_{x}\psi_{iR\uparrow}(x)-\psi_{iL\downarrow}^{\dagger}(x)\partial_{x}\psi_{iL\downarrow}(x)]dx $,

$ H_{i}= \sum\limits_{i=1,2}\sum\limits_{\alpha=R,L}\frac{g_{4}}{2}\int\psi_{i\alpha\sigma}^{\dagger}(x)\psi_{i\alpha\sigma}(x)\psi_{i\alpha\sigma}^{\dagger}(x)\psi_{i\alpha\sigma}(x)dx + g_{2}\int\psi_{iR\uparrow}^{\dagger}(x)\psi_{iR\uparrow}(x)\psi_{iL\downarrow}^{\dagger}(x)\psi_{iL\downarrow}(x)dx$

$ H_{d}=\sum\limits_{\sigma}\epsilon d_{\sigma}^{\dagger} d_{\sigma} +  U d_{\uparrow}^{\dagger} d_{\uparrow}d_{\downarrow}^{\dagger} d_{\downarrow} $, 

$ H_{t}=\sum\limits_{i=1,2}t(d_{\uparrow}^{\dagger}\psi_{iR\uparrow} + d_{\downarrow}^{\dagger}\psi_{iL\downarrow}) + h.c. $.

$H_{0}$ and $ H_{i}$ denote the kinetic energy and the interaction of the edge states respectively. $H_{d}$ is the Hamiltonian of the quantum dot where $\epsilon$ is the single particle energy level at the dot and $U$ is the on-site interaction energy. $ H_{t}$ describes the couplings between the dot and the edges and we have tuned to the channel symmetric point such that $t$ is independent of the edge index. In the presence of spin orbit coupling, spin ($\uparrow $ and $\downarrow$) denotes pseudo-spin of the edge modes. In the large $ U $ limit, one obtains the Kondo Hamiltonian $ H = H_{0} + H_{K} $ through perturbation theory,\cite{Hew} where 
\begin{equation}
H_{K}= \sum\limits_{i=1,2}J_{1}\vec{S}\cdot (\psi_{i\alpha}^{\dagger}\cdot\frac{\sigma_{\alpha\beta}}{2} \cdot \psi_{i\beta}) + \sum\limits_{i\ne j}J_{2}\vec{S} \cdot (\psi_{i\alpha}^{\dagger}\cdot\frac{\sigma_{\alpha\beta}}{2} \cdot \psi_{j\beta}).   \label{HK}
\end{equation}
Initially, $ J_{1}=J_{2} > 0 $ and without interaction, Eq.\ref{HK} reduces to a single channel Kondo problem.\cite{GR,NL} Our goal is to show that $J_{2} $ normalizes to zero at low temperatures when $g_{2}>0$. Because of helicity, one of the R/L and the spin $ \uparrow \downarrow $ indexes are redundant, we keep only one of them starting from Eq.\ref{HK}. 

It is important to note that, in general, there are extra backscattering terms of the form $ \psi_{iR}^{\dagger}(0)\psi_{iL}(0) + h.c. $ caused by the quantum dot in the Hamiltonian $ H $. The backscattering terms cut each edge into two separate parts at low temperatures if $K<1$ and render the geometry in Fig.1a unstable. However, in our case, single particle back-scattering terms are forbidden because of time-reversal symmetry. Two-particle backscatterings preserve time-reversal symmetry but they are irrelevant so long as $ K<1/4$.\cite{WBZ} Thus, our discussion below is valid for $ K > 1/4$.

By Abelian Bosonization,\cite{GNT,Gia} the electron operators can be written as: $\psi_{iR/L}=\frac{1}{\sqrt{2\pi a}}e^{\pm i(\sqrt{4\pi}\phi_{i R/L}(x)+k_{F}x)}$. We define Bosonic fields $ \phi_{i}(x)=\phi_{iL}(x) + \phi_{iR}(x)$ and $ \theta_{i}(x)=\phi_{iL}(x) - \phi_{iR}(x)$. The dual fields satisfy the commutation relations $ [\phi_{i}(x), \theta_{j}(x')] = \frac{-i}{2} \delta_{ij}\text{sgn}(x-x') $, where $\text{sgn}(x)=0$ when $x=0$. The symmetric and antisymmetric combinations of $ \phi_i $ and $ \theta_i $ are denoted as: $ \phi_{s/a}=\frac{1}{\sqrt{2}}(\phi_1 \pm \phi_2)$ and $ \theta_{s/a}=\frac{1}{\sqrt{2}}(\theta_1 \pm \theta_2)$. In terms of $ \phi_{s/a} $ and $ \theta_{s/a} $, the above Hamiltonians can be written as:
$H_{0}+ H_{i}= \frac{v'_{F}}{2}\int \frac{1}{K}(\partial_{x}\phi_{s})^{2}+ K(\partial_{x}\theta_{s})^{2} + \frac{1}{K}(\partial_{x}\phi_{a})^{2} + K(\partial_{x}\theta_{a})^{2}dx $ and 
$\begin{array}{l}
H_{K}=-\sqrt{\frac{2}{\pi}}J_{1}^{z}S_{z}\partial_{x}\theta_{s}(0) \\ + \frac{J_{1}^{xy}}{\pi a}(S^{-}e^{-i\sqrt{2\pi}\phi_{s}(0)}+ S^{+}e^{i\sqrt{2\pi}\phi_{s}(0)})\cos(\sqrt{2\pi}\phi_{a}(0))\\ + \frac{2J_{2}^{z}}{a \pi} S^{z}\sin(\sqrt{2\pi}\theta_{a}(0))\sin(\sqrt{2\pi}\phi_{a}(0)) \\+\frac{J_{2}^{xy}}{\pi a}(S^{-}e^{-i\sqrt{2\pi}\phi_{s}(0)}+ S^{+}e^{i\sqrt{2\pi}\phi_{s}(0)})\cos(\sqrt{2\pi}\theta_{a}(0)).
\end{array}$

Here, the Luttinger liquid parameter is defined as $ K =\sqrt{ \frac{1+{g_4}/{2\pi v_F} -{g_2}/{2 \pi v_F}}{1+{g_4}/{2\pi v_F} +{g_2}/{2 \pi v_F}}} $ and $ v'_{F} = v_{F}\sqrt{(1+{g_4}/{2\pi v_F})^2 - ({g_2}/{2 \pi v_F})^2} $.\cite{Gia} The Bosonic form of $H_{K}$ is spin anisotropic, we denote the two components of $ J_{i}$ as $ J_{i}^{z/xy}$ respectively. This is reasonable for our problem, because spin-orbit coupling breaks $SU(2)$ symmetry of the electrons on the edges. The spin-coupling terms acquire different scaling dimensions when $ K \neq 1 $ as we see below.

\emph{Renormalization Group Analysis}---With the Bosonic Hamiltonian $ H = H_{0} + H_{i} + H_{K} $, we may calculate the scaling dimensions of the operators in $ H_{K}$. At the vicinity of the fixed point where $ J_{i}^{z}=J_{i}^{xy}=0 $, the scaling dimensions of the $J_{1}^{z}$, $J_{1}^{xy}$ terms are $1$ and $ K $ respectively. On the other hand, the scaling dimensions of the $J_{2}^{z}$ and $J_{2}^{xy}$ terms are $ \frac{1}{2}(K+1/K)>1$. Thus, when $ K<1 $, the $J_2$ terms decrease but the $ J_{1} $ terms grow when temperature is lowered.

In order to study the physics at the strong coupling regime when $ J_{1} $ is of order 1, we use the Emery-Kivelson method by applying a unitary transformation $ U=e^{i\sqrt{2\pi K}\phi_{s}(0)S^{z}} $ to $ H $.\cite{EK} After the unitary transformation and the rescaling of the bosonic fields, we have:
\begin{equation}
\begin{array}{l}
\tilde{H}= U^{\dagger}HU \\ = H_{0}+ \lambda S^{z}\partial_{x}\theta_{s}(0)+ \frac{2J_{2}^{z}}{a \pi} S^{z}\sin(\sqrt{\frac{2\pi}{K}}\theta_{a}(0))\sin(\sqrt{2\pi K}\phi_{a}(0)) \\ +(S^{-}+ S^{+})[\frac{J_{1}^{xy}}{\pi a}\cos(\sqrt{2\pi K}\phi_{a}(0)) +\frac{J_{2}^{xy}}{\pi a}\cos(\sqrt{\frac{2\pi}{K}}\theta_{a}(0))],
\end{array}
\end{equation}
where $\lambda=\sqrt{2\pi K}v'_{F} - \sqrt{\frac{2}{\pi K}}J_{1}^{z}$. At the vicinity of the point $ \lambda=J_{1}^{xy}=J_{2}^{z}=J_{2}^{xy}=0 $, we may calculate the scaling dimensions of the $ J_{i} $ terms as before. The $ J_{1}^{xy} $ and $ J_{2}^{xy} $ terms become more relevant because of the elimination of the $ e^{\pm i \phi_{s}(0)}$ factors by the transformation. The $J_{1}^{xy}$ term has scaling dimension $ \frac{K}{2} $. It is relevant when $ K<1 $. The $J_{2}^{xy}$ term has scaling dimension $ \frac{1}{2K} $ and it is relevant when $ 1/2 < K < 1 $.

If $ \phi_{a} $ and $ \theta_{a} $ are independent fields, both $ J_{1}^{xy} $ and $ J_{2}^{xy} $ flow to infinite. As a result, the two-channel fixed point cannot be reached when $ 1/2 < K < 1$. That is the case for a quantum dot couples to two Luttinger liquid leads.\cite{FG} However, in the present problem, $ \phi_{a} $ and $ \theta_{a} $ are dual fields of each other. This fact changes the physics quite dramatically.

When $ K<1 $, the $ J_{1}^{xy}$ term is the most relevant term. At low temperatures, one expect the effective Hamiltonian of $ \tilde{H} $ is described by the fixed point Hamiltonian:
\begin{equation}
\tilde{H}_{eff}= H_{0}+ \frac{2J_{1}^{xy}}{\pi a}S^{x}\cos(\sqrt{2\pi K}\phi_{a}(0)). \label{Heffective}
\end{equation}
To show that this fixed point is indeed stable, we need to calculate the scaling dimensions of the operators $ J_{2}^{xy}S^{x}\cos (\sqrt{\frac{2 \pi}{K}}\theta_{a}(0)) $,  $ \delta\lambda S_{z}\partial_{x}\theta_{s}(0)$ and $ J_{2}^{z}S_{z}\sin(\sqrt{\frac{2\pi}{K}}\theta_{a}(0))\sin(\sqrt{2\pi K}\phi_{a}(0)) $ at the vicinity of the fixed point. $ \delta \lambda $ denotes the deviation from the Emery-Kivelson line. The results are given in the third column in Table I. We see that all these operators are irrelevant if $K<1$. Which confirms that Eq.\ref{Heffective} is the low temperature effective Hamiltonian. Below we give some details of the calculation.

To calculate the scaling dimension of $ J_{2}^{xy}S^{x}\cos (\frac{2 \pi}{K}\theta_{a}(0)) $, we observe that $ S^{x} $ commutes with $ \tilde{H}_{eff} $, so that we may set $ S^{x} $ to $ \pm 1/2 $. As a result, one needs to calculate the scaling dimension of $ \cos (\sqrt{\frac{2 \pi}{K}}\theta_{a}(0)) $ with the Hamiltonian 
\begin{equation}
\tilde{H}_{\pm}=H_{0} \pm \frac{J_{1}^{xy}}{\pi a}\cos(\sqrt{2\pi K}\phi_{a}(0)). \label{backscattering}
\end{equation}

Eq.\ref{backscattering} has the form of the Hamiltonian of a spinless Luttinger liquid wire with an impurity backscattering term at point $x=0$. One may interpret the $ \cos(\sqrt{2\pi K}\phi_{a}(0))$ as an impurity backscattering term with Luttinger liquid parameter $ K'=K/2 $. In a spinless Luttinger liquid, the backscattering term is relevant when $ K' < 1 $. At low temperatures, the ``backscattering term" $\cos(\sqrt{2\pi K}\phi_{a}(0))$ cuts the Luttinger liquid wire into two separate pieces at $ x=0 $.\cite{KF} As a result, the point $ x=0 $ is not a point at the bulk anymore, it can be regarded as a point at the boundary when $J_{1}^{xy} \to \infty $. In the bulk, the scaling dimension of $ \cos\sqrt{\frac{2\pi}{K}}\theta_{a}(0) $ is $ 1/2K $. However, at the boundary, $ \phi_{a}(0) $ is pinned to a constant value when $J_{1}^{xy} \to \infty $. Thus, the value of $ \phi_{a}(x) $ for $x \to 0 $ is constrained to be near the value of $\phi_{a}(0)$, by the $(\partial_x \phi_{a}(x))^{2}$ term in $ H_{0}$. Hence, the fluctuation of $ \theta_{a}(0) $ is enhanced. Detail calculations show that the scaling dimension of  $ \cos\sqrt{\frac{2\pi}{K}}\theta_{a}(0) $ should be $ 1/K $ instead.\cite{Gia} Consequently,  the  $ \cos\sqrt{\frac{2\pi}{K}}\theta_{a}(0) $ term is irrelevant when $ K <1 $.

The calculations of the scaling dimension of the $ S^{z}\partial_{x}\theta_{s}(0) $ term is similar to what have been done before in Ref.\onlinecite{FG2}. Since $\theta_{s}$ is governed by a free Hamiltonian, we have $<S^{z}(\tau)\partial_{x}\theta_{s}(\tau)S^{z}(0)\partial_{x}\theta_{s}(0)> = <S^{z}(\tau)S^{z}(0)><\partial_{x}\theta_{s}(\tau)\partial_{x}\theta_{s}(0)> $

In order to calculate $<S^{z}(\tau)S^{z}(0)>_{\tilde{H}_{eff}}$, we note that $ S^{z} $ is the sum of the raising and lowering operators of the eigenstates of $ S^{x} $, $|\pm>$. Thus, we have 
\begin{equation}
\begin{array}{l}
<-|e^{\tilde{H}_{eff}\tau}S^{z}(0)e^{-\tilde{H}_{eff}\tau}S^{z}(0)|-> \\
%=<-|e^{H_{-}\tau}e^{-H_{+}\tau}|-> \\
=<-|e^{\tilde{H}_{-}\tau}W^{\dagger}(0)e^{-\tilde{H}_{-}\tau}W(0)|->\\
=<-|W^{\dagger}(\tau)W(0)|->_{\tilde{H}_{-}}
\end{array}
\end{equation}
where $ W$ is a unitary transformation which satisfies $ W^{\dagger}e^{-\tilde{H}_{-}\tau}W = e^{-\tilde{H}_{+}\tau} $.

Actually, we need to find a unitary operator $ W $ which transforms $ \phi_{a} $ to $\phi_{a}+\sqrt{\frac{\pi}{2K}}$. Indeed, $ W=e^{-i\sqrt{\frac{\pi}{2K}}(\theta_{a}(L)-\theta_{a}(-L))} $, where $2L$ is the length of the system. In order to calculate $<-|W^{\dagger}(\tau)W(0)|->_{\tilde{H}_{-}}$, we come back to the spinless Luttinger liquid Hamiltonian interpretation of $ \tilde{H}_{-} $.

Suppose we have a spinless Luttinger liquid wire with an impurity at $x=0$ and the left and right ends locate at $-L$ and $ L $ respectively. At the strong coupling fixed point, the wire is cut into two separate parts by the impurity. However, the left half and the right half are described by  free Hamiltonians $ H_{0L} $ and $ H_{0R} $ respectively. As a result, we have
\begin{equation}
\begin{array}{l}
<W^{\dagger}(\tau)W(0)>_{\tilde{H}_{-}}\\
\approx <e^{i\sqrt{\frac{\pi}{2K}}(\theta_{a}(-L,\tau)-\theta(-L,0))}>_{H_{0L}}<e^{i\sqrt{\frac{\pi}{2K}}(\theta_{a}(L,\tau)-\theta(L,0))}>_{H_{0R}}\\
\propto \frac{1}{\tau^{\frac{1}{K}}}.
\end{array}                                                         
\end{equation}  
Since $ \partial_{x}\theta_{s}(0) $ has a scaling dimension of $1$, we conclude that the operator $ S^{z}\partial_{x}\theta_{s}(0) $ is irrelevant with scaling dimension $ 1/2K +1 $.

One may calculate the scaling dimension of the $ J_{2}^{z}S_{z}\sin(\sqrt{\frac{2\pi}{K}}\theta_{a}(0))\sin(\sqrt{2\pi K}\phi_{a}(0)) $ term in a similar way. This term has scaling dimension $ \frac{1}{2K} + \frac{1}{K} $. It is irrelevant when $ K<1$.

Up to now, we have neglected the $\sum\limits_{i\ne j}V_{2}\psi_{i\alpha}^{\dagger}\psi_{j\alpha} $ term in $ H_{K}$. However, this term is always irrelevant and we will drop this term for the rest of the paper.
\begin{table}
\caption{\label{T1} Scaling dimensions of the operators (Op.) at different fixed points (FPs). ($J_{2}=0$ for all fixed points, only coefficients of corresponding operators shown.)}
\begin{ruledtabular}
\begin{tabular}{|c|c|c|c|}
$\text{Op.} \setminus \text{FPs}$ & $ J_{1}=0 $ & $ \lambda=J_{1}^{xy}=0 $ & $ \lambda=0,  J_{1}^{xy} \to \infty $  \\ \hline
$ J_{1}^{z} (\delta \lambda) $ & $ 1 $ & $ 1 $ & $ 1+1/2K $ \\   \hline
$ J_{1}^{xy}$ & $ K $ & $ K/2 $ & $ N/A $ \\    \hline
$ J_{2}^{z} $ & $ \frac{1}{2}(K +1/K) $ & $ \frac{1}{2}(K +1/K) $ & $ 1/K +1/2K $ \\     \hline
$ J_{2}^{xy}$ & $ \frac{1}{2}(K +1/K)  $ & $ 1/2K $ & $ 1/K $ \\     \hline
$ V_{2} $ & $ \frac{1}{2}(K +1/K) $ & $ \frac{1}{2}(K +1/K) $ & $ 1/K $ \\     \hline
\end{tabular}
\end{ruledtabular}
\end{table}

\emph{Impurity Entropy}--- We have shown above that the two-channel fixed point is stable. At the vicinity of this fixed point, the system is described by $ \tilde{H}_{eff} $ in Eq.\ref{Heffective}. It is interesting to note that $ \tilde{H}_{eff} $ can be regarded as a generalization of the fixed point Hamiltonian of the two-channel and four-channel Kondo non-Fermi liquid fixed point Hamiltonians. The two-channel and four-channel cases are described by $ \tilde{H}_{eff} $ with $ K=1 $ and $ K=3/2 $ respectively.\cite{Comment} One of the most remarkable features of the multichannel Kondo effect is the existence of fractionally degenerate ground state. For example, the two-channel and four-channel Kondo models have residual entropy of $\frac{1}{2}\ln2 $ and $\frac{1}{2}\ln3 $ respectively.\cite{Tsv,AL}  This motivates us to study the residual entropy of $ \tilde{H}_{eff} $ for $ K<1$.

In order to calculate the residual entropy, we first calculate the partition function  $ Z= Tr \big\{ \exp^{-\beta \tilde{H}_{eff}} \big\} $. Since $ S^{x} $ commutes with the $  \tilde{H}_{eff} $, the states can be labeled as $ | \pm \frac{1}{2}, \phi_a> $.  We may write $ Z= Z_{+}+ Z_{-} = Tr \big\{ \exp^{-\beta \tilde{H}_{+}} \big\} + Tr \big\{ \exp^{-\beta \tilde{H}_{-}} \big\} $. Remember that the integral of $ H_{0} $ run from $ - \infty $ to $ + \infty $ if $ L \to \infty $ is taken. However, we may define new bosonic fields such that the positive axis is folded to the negative axis.\cite{FLS} After folding, the effect of the cosine term in $ \tilde{H}_{\pm}$ is to introduce a Dirichlet boundary condition. The actual value of $ \phi_{a} (0,\tau) $, whether it is pinned at $(2n+1)\pi/\sqrt{2\pi K} $ or $2n\pi/\sqrt{2\pi K} $, does not affect the partition function.\cite{Sal} Thus, we have $ Z_{+} =  Z_{-} $ and $ Z=2Z_{+}$. The impurity entropy of $\tilde{H}_{+}$ has been calculated in Ref.\onlinecite{FLS}. It is $ \ln g = \ln\sqrt{K/2} $, where g is the ``ground state degeneracy" discussed in Ref.\onlinecite{AL}. Together with the contribution from $S^{x}$, the total impurity entropy is 
\begin{equation}
S = \ln\sqrt{2K}.       \label{entropy}
\end{equation}
Substituting $ K=1 $ and $ K=3/2 $ into Eq.\ref{entropy} reproduces the impurity entropy results, calculated by the Bethe ansatz and boundary conformal field theory, for the two-channel and four-channel Kondo effects with non-interacting electrons respectively.\cite{Tsv,AL}

\emph{Specific Heat}---We show above that the residual impurity entropy is determined by the boundary term. On the other hand, the thermodynamic properties are determined by the leading irrelevant operators (LIO) near the fixed point. Since the irrelevant operators have correlation functions of the form $<<O(\tau)O(0)>> =[\frac{\pi T}{\sin(\pi T \tau)}]^{2\Delta}$, the second order correction of the free energy $\delta F^{(\lambda)}(T)= -\lambda^{2}  \int_{\tau_0}^{\beta/2} d\tau [\frac{\pi T}{\sin(\pi T \tau)}]^{2\Delta}$ where $\tau_0 $ is the cut-off time.\cite{FG2}

Near the strong coupling fixed point, the $\frac{J_{2}^{xy}}{\pi a}\cos(\sqrt{\frac{2\pi}{K}}\theta_{a}(0))$ term with scaling dimension $ 1/K $ is the LIO if $1/2< K <1$. We have $ \delta F \propto T^{2/K-1} + 0(T^2) $, where the $T^{2/K-1}$ term is universal and independent of the cut-off time $ \tau_0$ and the $0(T^2) $ term has a $\tau_0 $ dependent coefficient. As a result, the impurity specific heat $ C_{imp} \propto T^{2/K-2} $ if $ 2/3 < K <1$. However, when $ K < 2/3 $, the free energy is dominated by the $ 0(T^2) $ term and $ C_{imp} \propto T $.

\emph{Linear Conductance}---In the above calculations, we have assumed that the two helical edges have equal chemical potential. If a small bias is applied across the two edges, there should be a current flowing from one edge to the other. At temperatures much higher than the Kondo temperature, the linear conductance across the edges can be written as $ G \propto [2(J_{2}^{xy})^2 +(J_{2}^{z})^{2}] T^{K+1/K-2} $ by pertubation to second order. At low temperatures, the conductance can be obtained by substituting the normalized value of $ J_{2} $ into the high temperature expression of $ G $. As we show above, $J_{2}$ decreases near the fixed point with $ J_{1}^{xy}=J_{1}^{z}=J_{2}=0 $ and $J_{1}$ grows. If $ 1/2<K<1$, however, when the $J_{1} $ terms are of order 1, the $ J_{2}^{xy} $ term becomes relevant and grows. When the temperature decreases further, the strong coupling fixed point with $ J_{1}^{xy} \to \infty $ is reached. Near this fixed point, $J_{2}^{xy} $ and $ J_{2}^{z} $ are irrelevant and flow to zero. As a result, the conductance acquires a non-trivial temperature dependence. On the other hand, if $ K<1/2 $, the $J_{2}$ terms are always irrelevant and decrease to zero. Consequently, the conductance decreases monotonically as the temperature lowers. The temperature dependence of the linear conductance at different fixed points are shown in the Fig.1b.

\emph{Conclusion}--- We show above that a quantum dot couples to two helical edge states of 2D topological insulators is described by a Kondo Hamiltonian. Weak repulsive interaction with $ K<1 $ on the edges, drives the system to a two-channel fixed point at low temperatures. At the fixed point, the residual impurity entropy is $ \ln\sqrt{2K} $. Near the fixed point, the impurity specific heat is $C_{imp} \propto T^{2/K-2} $ when $ 2/3<K<1 $ and $C_{imp} \propto T $ when $K<2/3$. Moreover, the linear conductance across the two edges through the quantum dot has non-trivial temperature dependence.

\emph{Acknowledgments}---
We thank Y. Avishai, C.H. Chung, D. Feldman and especially C.L. Kane for inspiring discussion and useful suggestions. KTL is supported by the IAS-HKUST postdoc fellowship, PAL acknowledge the support of NSF DMR0804040 and the hospitality of IAS-HKUST.

\end{document}